\documentclass[traditabstract,twocolumn]{aa}
\usepackage{natbib}
\usepackage{graphicx}
\usepackage{txfonts}
\usepackage{mathrsfs}
\usepackage{mathbbol}
\DeclareMathAlphabet{\mathpzc}{OT1}{pzc}{m}{it}
\begin{document}

%

\def\kms{\,{\rm km}\,{\rm s}^{-1}}
\def\hmpc{\,{h^{-1} {\rm Mpc}}}
\def\mpch{\,{{\rm Mpc}^{-1} {\rm h}}}
\def\galaxy{{\tt GALAXY\,}}
\def\group{{\tt GROUP\,}}
\def \omgmh{{$\Omega^{0.6}_{\rm m}h$\,}}
\def \cf2{{\it Cosmicflows-2\,}}
\def \wmap{{\it WMAP\,}}
\def \planck{{\it Planck\,}}
\def \der{{\rm d}}




\setcounter{MaxMatrixCols}{10}


\title
{Constraining cosmology with pairwise velocity estimator}
\author
{Yin-Zhe Ma$^{1,2\dagger}$, Min Li$^{3,\ddagger}$, \& Ping He$^{4,5,\star}$}
\offprints{Y.-Z.~Ma \email{ma@ukzn.ac.za}}
\authorrunning{Y.-Z.~Ma, M.~Li, P.~He}
\titlerunning{Pairwise velocity estimator}
\institute{$^1$Jodrell Bank Centre for Astrophysics, School of Physics and Astronomy, The University of Manchester, Manchester M13 9PL,
UK\\
$^{2}$School of Chemistry and Physics, University of KwaZulu-Natal, Westville Campus, Private Bag X54001, Durban, 4000, South Africa \\
$^3$Changchun Observatory, National Astronomical Observatories, CAS, Changchun, Jilin 130117, China\\
$^4$College of Physics, Jilin University, Changchun 130012, China\\
$^5$Center for High Energy Physics, Peking University, Beijing 100871, China} 

\abstract{In this paper, we develop a full statistical method for the pairwise velocity estimator previously proposed, and apply \cf2 catalogue to this method to constrain cosmology. We first calculate the covariance matrix for line-of-sight velocities for a given catalogue, and then simulate the mock full-sky surveys from it, and then calculate the variance for the pairwise velocity field. By applying the $8315$ independent galaxy samples and compressed $5224$ group samples from \cf2 catalogue to this statistical method, we find that the joint constraint on $\Omega^{0.6}_{\rm m}h$ and $\sigma_{8}$ is completely consistent with the {\it WMAP} 9-year and {\it Planck} 2015 best-fitting cosmology. Currently, there is no evidence for the modified gravity models or any dynamic dark energy models from this practice, and the error-bars need to be reduced in order to provide any concrete evidence against/to support $\Lambda$CDM cosmology.}



\vskip 0.3 truein

\maketitle


\section{Introduction}
\label{sec:intro}

The study of peculiar velocity field is a powerful tool for probing the large-scale structures of the Universe. This is because in the standard $\Lambda$ cold dark matter ($\Lambda$CDM) cosmology, the gravitational instability causes the density perturbations to grow and the peculiar velocity field to emerge. On large scales, the peculiar velocity field is directly related to the underlying matter perturbations that can be used to test the growth of structure in the standard $\Lambda$CDM universe.

There have been various approaches to using the peculiar velocity field to study cosmology and probe the growth of structure. Since in the linear perturbation theory, the velocity field in real space is related to the integral of matter density contrast with a Newtonian kernel \citep{Peebles93}, there has been wide interest in comparing the measured velocity field with the reconstructed density field from galaxy redshift surveys and testing the linear relation between the two. One of the approaches is to reconstruct the linear velocity field from the density field and compare it with the measured velocity field~\citep{Branchini01,Pike05,Davis11,Ma12}. The other approach, the ``POTENT'', is to use the reverse process, i.e. reconstructing the gravitational potential and density field from the velocity field and comparing with the measured galaxy density field~\citep{Dekel93,Dekel99,Hudson95,Sigad98,Branchini00}. The results from these practices show that the linear perturbation theory works very well on scales of $10$--$100\hmpc$, and the fitted growth rate factors ($f\sigma_{8}$) are consistent with the $\Lambda$CDM cosmology at low redshifts. The second method is to reconstruct the cosmic bulk flow on various depths of the local Universe, which are only sensitive to cosmological perturbations on large scales~\citep{Kashlinsky08,Watkins09,Feldman10,Watkins15}. In recent years, there have been a few studies that claim to find very large bulk flows on a scale of $100\hmpc$ or on deeper scales that seem to exceed the $\Lambda$CDM prediction by a $3\sigma$ confidence level (CL)~\citep{Kashlinsky08,Watkins09,Feldman10,Macaulay11,Macaulay12}. But later studies show that this might be due to the systematics that arose when combining different catalogues with different calibration schemes~\citep{Nusser11,Ma13}. The third method is to directly fit the velocity field power spectrum from the peculiar velocity field data~\citep{Macaulay11,Macaulay12,Johnson14}. The recent results from the six-degree-field galaxy survey data (6dF) show that the fitted values of structure growth rate at low redshift are consistent with {\it Planck} 2013 cosmology~\citep{Planck2013-16}.

In this paper, we consider a different estimator of peculiar velocity field, namely the mean relative pairwise velocity of galaxies $v_{12}$, which is defined as the mean value of the peculiar velocity difference of a galaxy pair at separation $r$~\citep{Ferreira99}. In the fluid limit, the pairwise velocity becomes a density-weighted relative velocity~\citep{Juszkiewicz98},
\begin{eqnarray}
\mathbf{v}_{12}(r)=\langle \mathbf{v}_{1}-\mathbf{v}_{2} \rangle_{\rho}=\frac{\langle (\mathbf{v}_{1}-\mathbf{v}_{2})(1+\delta_{1})(1+\delta_{2}) \rangle}{1+\xi(r)}, \label{eq:pairwise-eq}
\end{eqnarray}
where $\mathbf{v}$ and $\delta$ are the peculiar velocity and density contrast, respectively, and $\xi$ is the two-point correlation function. Since the line-of-sight velocities of discrete galaxies are measured for each sample, \citet{Ferreira99} proposed that the estimator of the pairwise velocity is
\begin{eqnarray}
v_{12}(r)=\left[2 \sum (s_{A}-s_{B})p_{AB}\right] \Big/ \left[\sum p^{2}_{AB}\right], \label{eq:pairwise-estimate}
\end{eqnarray}
where $s_{A}=\mathbf{v}_{A}\cdot \hat{\mathbf{r}}_{A}$ is the line-of-sight velocity of galaxy A, $p_{AB}=\hat{\mathbf{r}}\cdot(\hat{\mathbf{r}}_{A}+ \hat{\mathbf{r}}_{B})$ is the geometric factor, and the summation is for all pairs within a distance separation bin $\Delta r$. It was first proposed in \citet{Ferreira99} that the estimator measures the cosmological density parameter $\Omega$, and later it was found in \citet{Juszkiewicz00} that the matter density of the Universe is close to $0.35$ and that the Einstein-de Sitter model $\Omega=1$ is inconsistent with the data. The result of a low-density universe was further confirmed by \citet{Feldman03} with the Mark III catalogue, Spiral Field I-Band (SFI) catalogue, Nearby Early-type Galaxies Survey (ENEAR) catalogue, and the Revised Flat Galaxy Catalog (RFGC). This becomes the early measurement of matter content of the Universe before the experiment of the cosmic microwave background radiation from {\it Wilkinson Microwave Anisotropy Probe} (\wmap) and strongly indicates that there is $\Lambda$ in the cosmic budget. In addition, \citet{Juszkiewicz99} investigated the dynamics of the pairwise motion by calibrating the mean pairwise velocity with N-body simulations. They provide a theoretical formula that is very consistent with the N-body simulation result. Additionally, \citet{Bhatt07} forecast the prospective constraints on cosmological parameters from the pairwise kinetic Sunyaev-Zeldovich effect (proportional to velocity field). In 2012, by applying the pairwise momentum estimator of~\citet{Ferreira99} into the temperature map, the Atacama Cosmology Telescope team (ACT) provided the first detection of the kinetic Sunyaev-Zeldovich effect~\citep{Hand12}. More recently, \citet{Planck2015-kSZ} have estimated the pairwise momentum of the kSZ temperature fluctuations of {\it Planck} maps at the positions of the Central Galaxy Catalogue samples extracted from Sloan Digital Sky Survey (SDSS-DR7) data. They find a $\sim 2\sigma$ CL detection of the kSZ signal, which is consistent and slightly lower than the one found in~\citet{Hand12}.

The literature listed above about the pairwise velocity estimator is for the old data set, whereas this work uses a new data set, the new compiled \cf2 catalogue, to constrain cosmology. In addition, we developed a new method of computing the covariance matrix of the pairwise velocity field and formed a likelihood that directly relates the models of pairwise velocity field with the data. The paper is organized as follows. In Sect.~\ref{sec:data}, we introduce the new \cf2 data set. In Sect.~\ref{sec:method}, we introduce our statistical method of computing the likelihood of pairwise velocities. In Sect.~\ref{sec:results}, we present our results of constraints and compare them with \wmap nine-year cosmology and \planck 2015 cosmology. The conclusion is presented in the last section.

Throughout the paper, unless otherwise stated, we use \planck 2015 best-fitting cosmological parameters~\citep{Planck2015-13}, i.e. \{$\Omega_{\rm m}$, $\sigma_{8}$, $h$, $\Omega_{\rm b}$, $\ln(10^{10}A_{\rm s})$\}=\{0.309, 0.816, 0.677, 0.049, 3.064\}.

\section{Data set}
\label{sec:data}
\begin{figure}
\centerline{\includegraphics[width=3.3in]{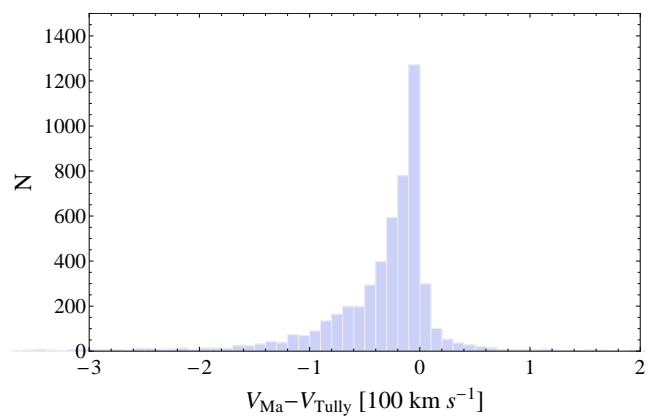}}
\caption{Histogram of the line-of-sight peculiar velocity difference in the \group catalogue between the calculation by using Eq.~(\ref{eq:vp}) (labelled as $V_{\rm Ma}$) and the values given in \citet{Tully13} (labelled as $V_{\rm Tully}$).} \label{fig:vdiff}
\end{figure}

\begin{figure*}
\centerline{\includegraphics[width=3.3in]{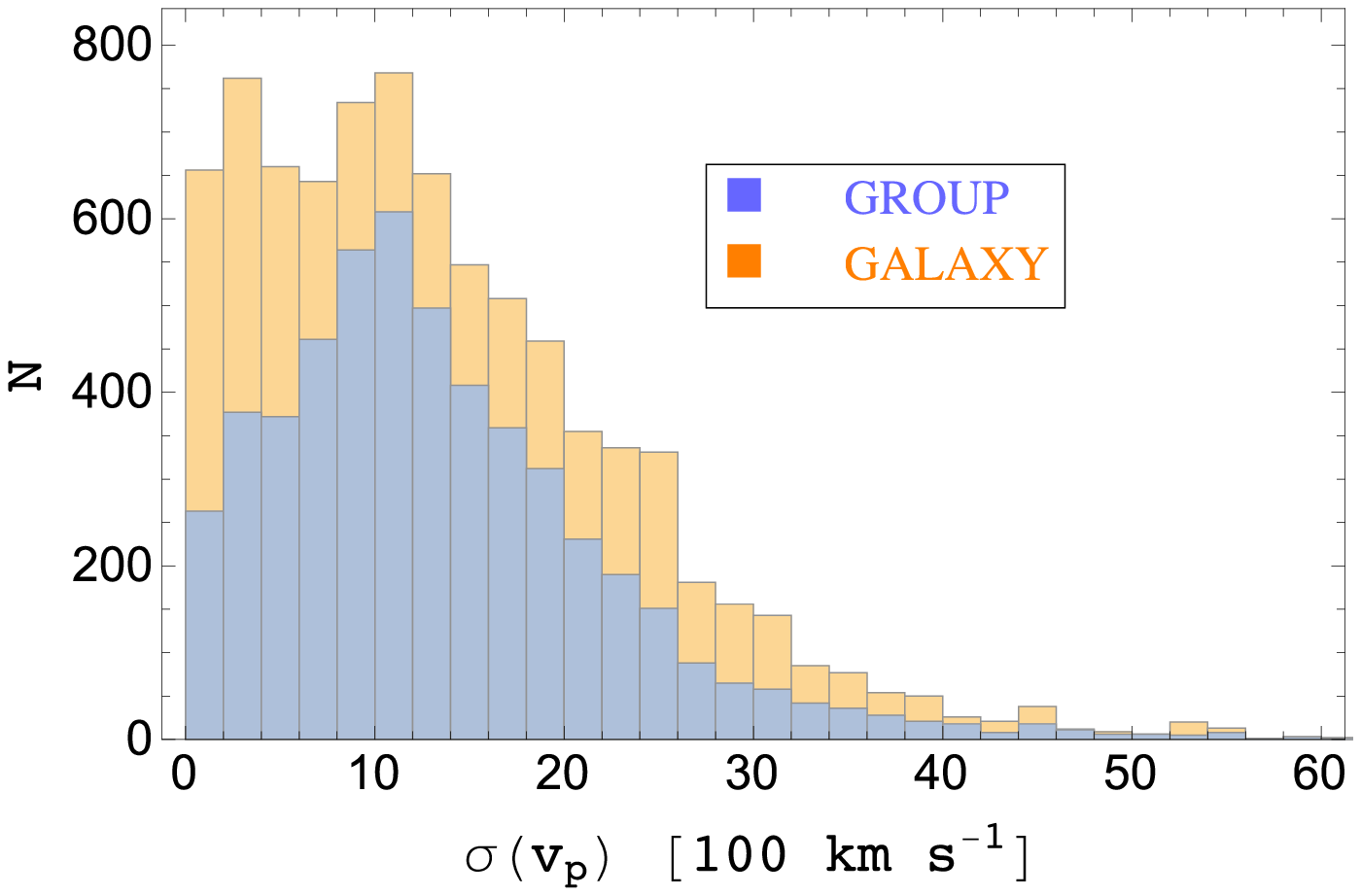}
\includegraphics[width=3.3in]{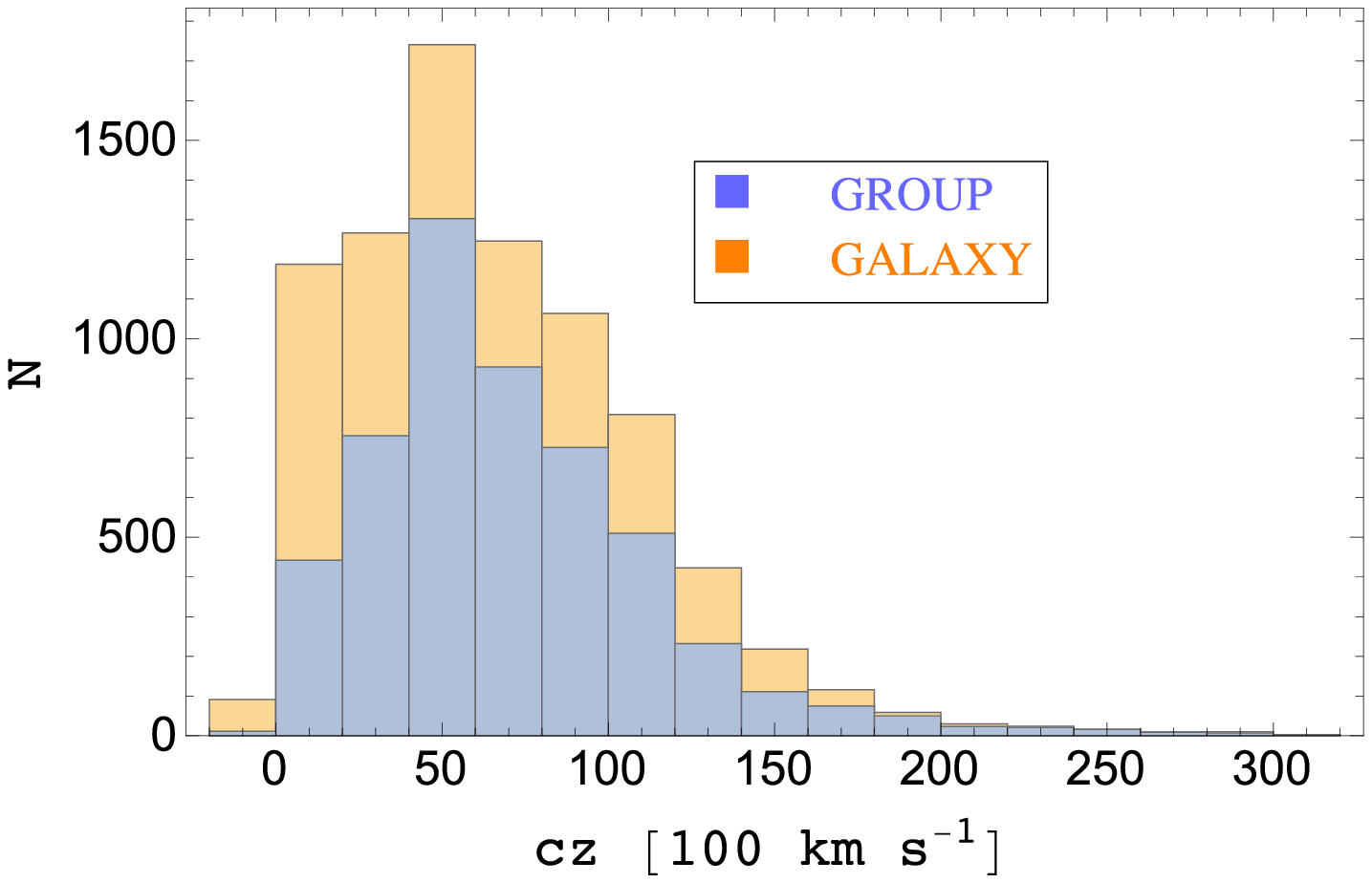}}
\caption{{\it Left}--Histogram of the measured velocity error of {\tt GALAXT} (Orange) and {\tt GROUP} (blue) catalogues. The two histograms peak at $500$--$1\ 000$ and $1\ 000\,{\rm km\,s^{-1}}$, respectively. The bin width of the histogram is $\Delta \sigma(v_{\rm p})=200\,{\rm km\,s^{-1}}$. {\it Right}--Histogram of redshift distribution of \galaxy and \group samples. The bin width of the plot is $c\Delta z= 2\,000\kms$.} \label{fig:vel-err}
\end{figure*}


\begin{figure*}
\centerline{\includegraphics[width=3.3in]{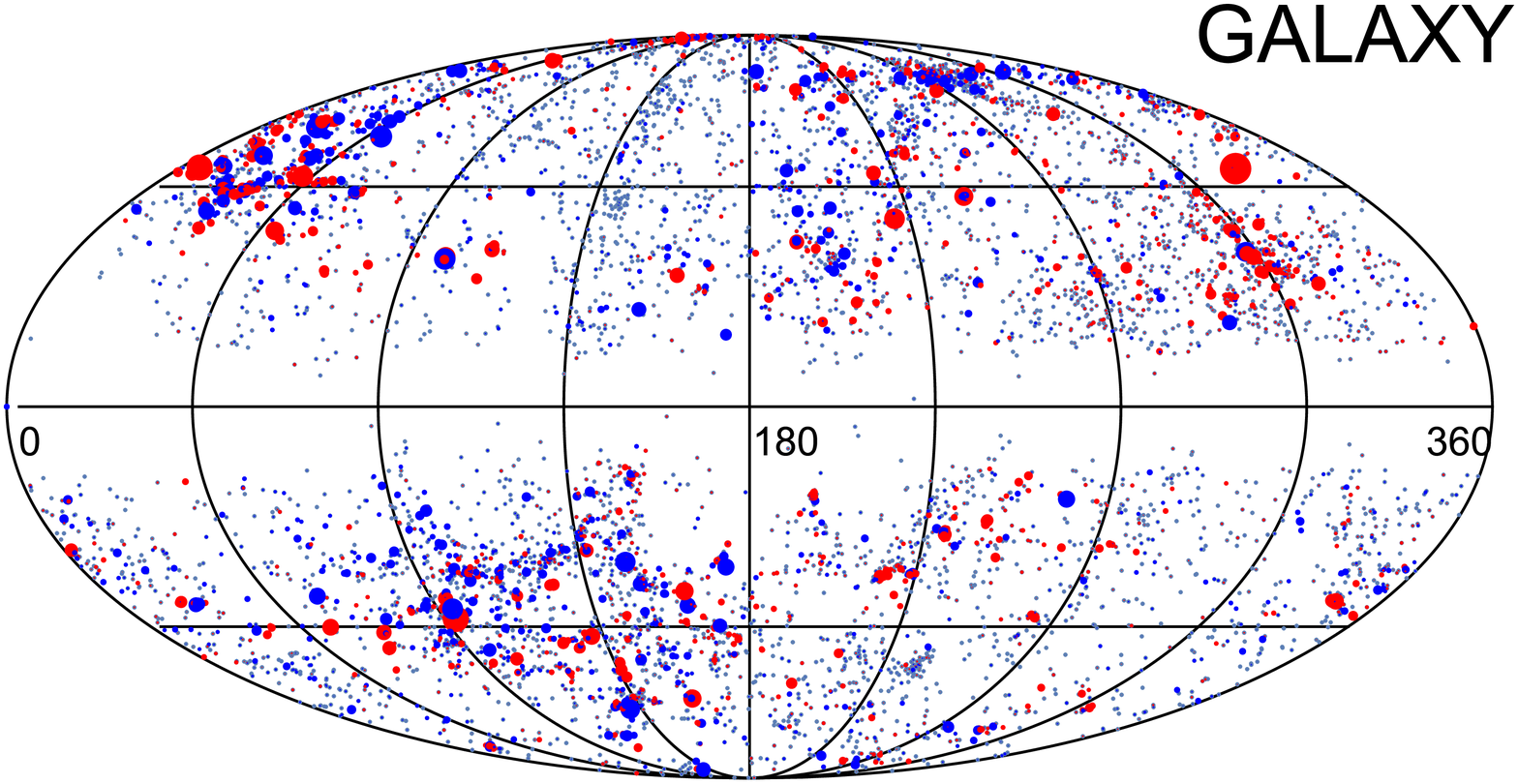}
\includegraphics[width=3.3in]{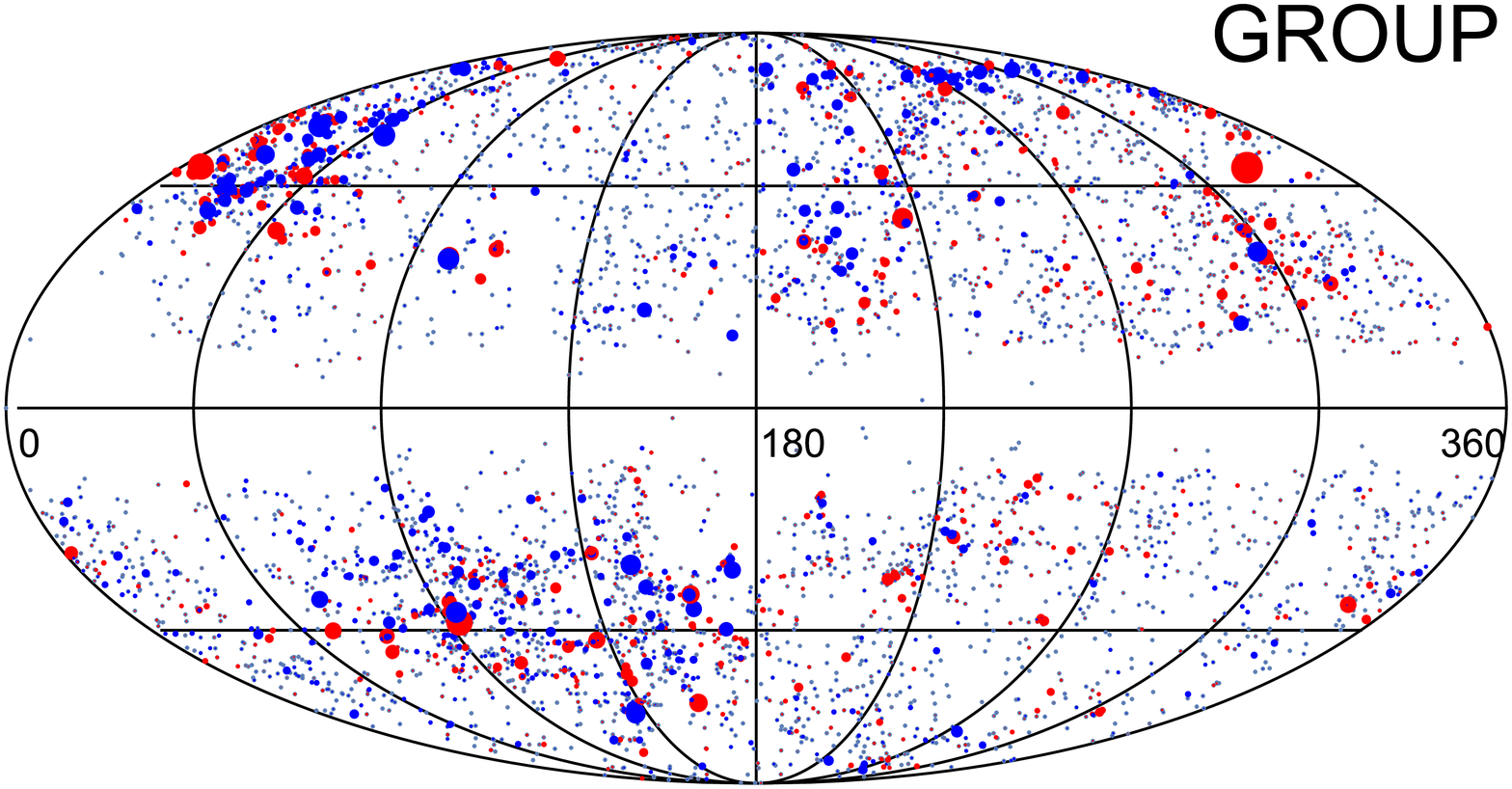}}
\caption{Full-sky \galaxy (8315 samples) and \group (5224 samples) catalogues plotted in Galactic coordinates. The red (light grey) points are moving away from us, and the blue (dark grey) points towards us. The size of the points is proportional to the magnitude of the line-of-sight peculiar velocity.} \label{fig:fullsky}
\end{figure*}

The \cf2 catalogue~\citep{Tully13} is a compiled catalogue of distances and peculiar velocities of more than $8\ 000$ galaxy samples. Some of these samples are from new measurements, while others are taken from the literature. The majority of the distances of the samples are measured through the Tully-Fisher relation~\citep{Tully77} or fundamental-plane (FP) relation~\citep{Djorgovski87,Campbell14} with roughly $20$ per cent of the error for the distances. But there are a small portion of the samples whose distances are measured from Type-Ia supernovae light curves, surface brightness fluctuation, the tip of the red giant
branch (TRGB), or Cepheids. The \cf2 samples are calibrated at their zero points by using two different approaches. One is to use the Cepheid period-luminosity relation (Cepheid PLR), and the other to use the luminosities of red giant branch stars at the onset of core helium burning, at a location in a stellar colour-magnitude diagram (CMD) known as the TRGB.

Eventually, the correlation between the galaxy luminosity and HI line width was refitted so zero points were determined~\citep{Tully13}. In addition, \citet{Tully12} have demonstrated that no extra systematics were found after the zero-point calibration. The results of the two methods of re-calibrations were compared, which confirmed that the distance estimation is unbiased~\citep{Tully13}. We obtained these samples from the ``VizieR'' astronomy data base, which are available in two tables. Table I provides the entry for every galaxy with a distance, consisting of 8315 galaxies in total. Table II condenses the galaxies in each group and provides a distance for each group, therefore consisting 5224 group entries. In the following analysis, we use both tables in the pairwise velocity analysis, so we name Table I as \galaxy and Table II as \group in the following text.

We made one adjustment of the observed peculiar velocities in the \cf2 samples. On page 20 of \citet{Tully13}, it is stated that the line-of-sight peculiar velocity is calculated through $cz-H_{0}d$, where $d$ is the measured distance. However, as shown in \citet{Davis14} and \citet{Watkins14}, this estimator makes the velocity estimate biased at the level $\Delta v_{\rm p}\sim 100 \kms$ at $z \sim 0.04$ and even more biased at higher redshifts. The more accurate formula for calculating the line-of-sight velocity is~\citep{Davis14}
\begin{eqnarray}
v_{\rm p}=c\left(\frac{z-\overline{z}}{1+\overline{z}}  \right), \label{eq:vp}
\end{eqnarray}
where $z$ is the measured redshift, and $\overline{z}$ is the redshift for the unperturbed background. One can obtain $\overline{z}$ by inverting the measured luminosity distance
\begin{eqnarray}
D_{\rm L}(\overline{z})=(1+\overline{z})\int^{\overline{z}}_{0}\frac{c\, \der z'}{H(z')}, \label{eq:luminosity}
\end{eqnarray}
where $H(z)=H_0\sqrt{\Omega_{\rm m}(1+z)^{3}+\Omega_{\Lambda}}$ is the Hubble parameter. In Fig.~\ref{fig:vdiff}, we show the measured velocity difference of \group catalogue between Eq.~(\ref{eq:vp}) and the Col.~20 of Table II in \citet{Tully13}. One can see that by adjusting the peculiar velocity calculation, there is a trend towards lower velocity values, which indicates that the original velocities provided by~\citet{Tully13} are biased towards higher velocity values. This is consistent with the prediction shown in Figs.~2 and 3 in \citet{Davis14}\footnote{The re-adjusted peculiar velocities of \cf2 catalogue can be requested from the corresponding author.}. In addition, we propagate the distance error into the peculiar velocities and plot the histogram of the {\it rms} velocity errors in the upper panel of Fig.~\ref{fig:vel-err}. One can see that the two histograms peak at slightly different values, i.e. $500$--$1\ 000$ and $1\ 000\,{\rm km\,s^{-1}}$, respectively. We use the propagated velocity errors as the measurement error in the likelihood analysis.


We plot the redshift distribution of \galaxy and \group catalogues in the lower panel of Fig.~\ref{fig:vel-err}. One can see that the bulk of the samples are in the range of $0$--$15\ 000\kms$, while a few samples reside at higher redshifts. The two catalogues are probing the same redshift range. The \group catalogue is just a scaled-down version of the \galaxy catalogue, since \group is just a condensation of the \galaxy catalogue. In Fig.~\ref{fig:fullsky}, we plot the re-adjusted peculiar velocities (through Eq.~(\ref{eq:vp})) of two catalogues on the sky in the Galactic coordinate, so the Galactic plane region is empty. The red (light grey) points are moving away from us, and the blue (dark grey) points are moving towards us. The size of the points is proportional to the magnitude of the line-of-sight peculiar velocity. One can see that the velocity distribution is almost uniform across the full sky.

\section{Methodology}
\label{sec:method}

\begin{figure}
\centerline{\includegraphics[width=3.3in]{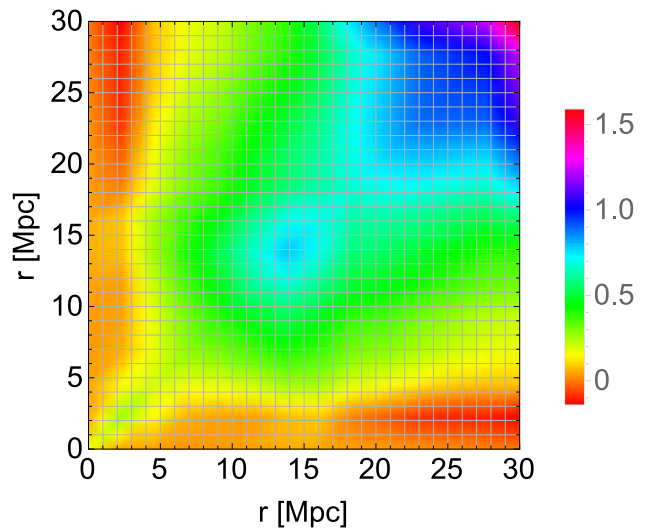}}
\caption{Covariance matrix of $v_{12}(r)$ for \galaxy catalogue. The $x-$ and $y-$ axes mean the different radial bins, and the unit of colour bar is $100\,{\rm km\,s^{-1}}$.} \label{fig:covfig}
\end{figure}

In this section, we first present three theoretical models of pairwise velocity field, and then discuss how can we simulate the covariance matrix of the pairwise velocity estimator at different distance bins. Then we present the likelihood function.

\subsection{Theoretical models}
\label{sec:models}
Equation~(\ref{eq:pairwise-eq}) is the definition of pairwise velocity field at each separate distance $r$. The approximate solution of the pairwise velocity field through the pair conservation equation derived by \citet{Juszkiewicz99} is given as (also see~\citealt{Feldman03})
\begin{eqnarray}
v_{12}(r) & = & -\frac{2}{3}\Omega^{0.6}_{\rm m} H_{0} r \bar{\xi}(r)(1+\alpha \bar{\xi}(r)), \\ \label{eq:v12-r-theory}
\bar{\xi}(r) & = & \left[3 \int^{r}_{0}\xi(x)x^{2}\der x\right] \Big/ \left[r^{3}(1+\xi(r))\right] ,\label{eq:xi-r}
\end{eqnarray}
where $\alpha=1.2-0.65\gamma$, and $\gamma=-(\der \ln \xi/\der \ln r)|_{\xi=1}$. $\xi(r)$ can be measured directly from galaxy surveys two-point correlation statistics. Here we use the
two-point correlation function of Point Source Catalogue redshift (PSC$z$ survey) calculated by \citet{Hamilton02}. This is because the PSC$z$ survey measures the galaxy distance out to $120\hmpc$ and its sky coverage is close to the \cf2 catalogue (see panels~a, b, and d of fig.~1 in \citet{Ma12}). Thus the PSC$z$ covers the similar volume of the local Universe as \cf2. The $\xi(r)$ model derived by PSC$z$ survey \citep{Hamilton02} is
\begin{eqnarray}
\xi(r)=\left(\frac{\sigma_{8}}{0.83} \right)^{2}\left[\left(\frac{r}{r_{1}} \right)^{-\gamma_{1}}  + \left(\frac{r}{r_{2}} \right)^{-\gamma_{2}} \right], \label{eq:model1}
\end{eqnarray}
where $r_{1}=2.33\hmpc$, $r_{2}= 3.51\hmpc$, $\gamma_{1}=1.72$,
$\gamma_{2}=1.28$, and $\sigma_{8}$ is a free parameter.

We name the above PSC$z$ model as Model 1 in the following discussion. In addition, we also consider two other simplified models of the correlation function as~\citep{Feldman03}
\begin{eqnarray}
\xi(r)=\left(\frac{\sigma_{8}}{0.83} \right)^{2}\left(\frac{r}{r_{0}} \right)^{-\gamma_{0}}, \label{eq:model23}
\end{eqnarray}
where ($\gamma_{0}$,$r_{0}$)=($1.3$, $4.76\hmpc$) and ($1.8$, $4.6\hmpc$) for Models 2 and 3, respectively.

We vary two parameters in the pairwise velocity field model. One is $\sigma_{8}$, which is the square root of the amplitude of the two-point correlation function (Eqs.~(\ref{eq:model1}) and (\ref{eq:model23})). But because $\xi(r)$ affects both the numerator and denominator of $\bar{\xi}(r)$ (Eq.~(\ref{eq:xi-r})), $\sigma_{8}$ will affect the shape of the pairwise velocity field function $v_{12}(r)$. In addition, we also vary the total amplitude parameter $\Omega_{\rm m}^{0.6}h$ as the pre-factor in Eq.~(\ref{eq:v12-r-theory}).

The three model predictions plotted in Fig.~\ref{fig:v12fig} are all negative in the regime of $0$--$30$ Mpc, because of the gravitational potential effect. In addition, one can see that Models 1 and 2 predictions are close to each other, while Model 3 has slightly lower power on a large separation $r\sim 30\,$Mpc.

\subsection{Covariance matrix}
\label{sec:covariance}
We now calculate the covariance matrix for the pairwise velocity estimator. In the original work of \citet{Ferreira99}, the covariance matrix is calculated analytically by assuming a smoothed survey window function. However, this is only an approximation since the real survey window can have complicated geometry. Here we calculate the covariance matrix of pairwise velocity field directly from a numerical simulation that automatically includes the real survey geometry.

We simulated a large number of mock catalogues. For each simulation, we simulated $N_{\rm data}=8315$ and $5224$ samples' line-of-sight velocities for \galaxy and \group catalogues and then computed the pairwise velocities from Eq.~(\ref{eq:pairwise-eq}). Then we calculated the variance of the pairwise velocity field from these mock catalogues. We tested that $N_{\rm sim}=10^{3}$ is large enough for the results to converge.

\subsubsection{Covariance matrix of line-of-sight velocities}

As one can see from Eqs.(14)--(16) of \citet{Ma13}, the line-of-sight velocities are correlated. We therefore need to calculate the covariance matrix of these line-of-sight velocities before simulation. For any two samples, their covariance matrix is
\begin{eqnarray}
G_{ij} &=& \langle v_{i}v_{j} \rangle +
\delta_{ij}(\sigma^{2}_{\ast}+\sigma^{2}_{i})  \notag \\
&=& \langle (\hat{\mathbf{r}}_{i} \cdot
\mathbf{v}(\mathbf{r}_{i}))
 (\hat{\mathbf{r}}_{j} \cdot \mathbf{v}(\mathbf{r}_{j})) \rangle
 +\delta_{ij}(\sigma^{2}_{i}+\sigma^{2}_{\ast}) ,  \label{vel_Gnm1}
\end{eqnarray}
where the first term is the cosmic variance term since the intrinsic correlation between velocities at two different directions. The second bracket contains the {\it rms} measurement noise for each galaxy $\sigma_{i}$ (i.e. quantity plotted in the upper panel of Fig.~\ref{fig:vel-err}) and the small scale and intrinsic dispersion $\sigma_{\ast}$ which is estimated to be around $200\kms$~\citep{Turnbull12}.


For the cosmic
variance term, any two samples are correlated so the off-diagonal element of $G_{ij}$ ($i\neq j$) is non-zero. But for
the measurement noise term, small scale, and intrinsic dispersion, the two samples are not correlated, so they only contribute to the diagonal term in the $G_{ij}$ matrix.
The first term
is the real space velocity correlation function, which is related
to the matter power spectrum in Fourier space \citep{Watkins09,Ma13},
\begin{equation}
\bigl\langle \bigl(\hat{\mathbf{r}}_{i} \cdot
\mathbf{v}(\mathbf{r}_{i})\bigr)
 \bigl(\hat{\mathbf{r}}_{j} \cdot \mathbf{v}(\mathbf{r}_{j})\bigr) \bigr\rangle
 =\frac{H^{2}_{0}f^{2}(z=0)}{2 \pi^{2}}\int \mathrm{\ } \der k
 \mathrm{\ } P(k) \mathrm{\ } F_{ij}(k),  \label{vel_rnrm1}
\end{equation}
where $P(k)$ is the matter power spectrum that we output from public code~{\sc camb}~\citep{Lewis00}. The $f(z)=\der \ln D/\der \ln a$ is the growth rate function that characterizes how fast the structures grow at different epochs of the Universe. Since the \cf2 samples peak at the redshift $z\simeq 0.0167$ (lower panel of Fig.~\ref{fig:vel-err}), we use the zero-redshift growth function $f(z=0)=\Omega_{\rm m}^{0.55}$~\citep{Watkins09,Ma13} in Eq.~(\ref{vel_rnrm1}). In the future, if the survey probes deeper region of the space, one should use the corresponding growth function $f(z)$ in Eq.~(\ref{vel_rnrm1}), so that the joint constraints on $f(z)\sigma_{8}$ can be obtained, which constitutes a sensitive test of modify gravity models~\citep{Hudson12,Planck2015-13}.

The $F_{ij}(k)$ is the window function
\begin{equation}
 F_{ij}(k)=\int \mathrm{\ }\frac{\der ^{2} \hat{k}}{4 \pi}
  \left(\hat{\mathbf{r}}_{i}\cdot \hat{\mathbf{k}} \right)
  \left(\hat{\mathbf{r}}_{j}\cdot \hat{\mathbf{k}} \right)
  \times \exp(i k \hat{\mathbf{k}}\cdot (\mathbf{r}_{i}
  -\mathbf{r}_{j})),  \label{vel_fmn_win}
\end{equation}
which can be calculated analytically (The Appendix in \citealt{Ma11}).

Therefore, by calculating $G_{ij}$ matrix, we obtain a $N_{\rm data}\times N_{\rm data}$ covariance matrix for the line-of-sight velocities of mock galaxies. We then followed the procedure in Appendix~\ref{sec:simulate} to simulate a mock line-of-sight velocity catalogue. We did this repeatedly for $N_{\rm sim}$ number of mock catalogues.

\subsubsection{Covariance matrix for pairwise velocities}
\label{sec:covariance-v12}
For each mock catalogue, we can plug them into Eq.~(\ref{eq:pairwise-estimate}) to obtain $v_{12}(r)$ for different bins, then one can obtain $N_{\rm sim}$ numbers of mock $v_{12}$ velocity fields. Then, for each distance bin, one can calculate the covariance matrix for $N_{\rm bin}$ as
\begin{eqnarray}
C_{ij} &=& \langle v_{12}(r_{i})
v_{12}(r_{j})  \rangle \nonumber \\
&= & \frac{1}{N_{\rm sim}} \sum^{N_{\rm sim}}_{k=1} \left(v^{k}_{12}(r_{i}) \right) \left(
v^{k}_{12}(r_{j})  \right) . \label{eq:cov-Cij}
\end{eqnarray}
Therefore, this $C_{ij}$ is a $N_{\rm bin} \times N_{\rm bin}$ positive-definite symmetric matrix which is the covariance matrix for $v_{12}(r)$.

In Fig.~\ref{fig:covfig}, we plot the covariance matrix of the \galaxy catalogue. One can see that it is a positive and symmetric matrix, and the higher value of correlation exists on large separation distances. This is because the cosmic variance term becomes more significant at a larger separation distance. In Fig.~\ref{fig:v12fig}, we plot the square root of the diagonal value of covariance matrix as the error bars for \galaxy and the \group catalogue as an example for $N_{\rm bin}=14$. But here we remind the reader that the correlation between different data points is significant, so the value of the square-root cannot fully represent the total error budget.

\subsection{Likelihood}
We now have the models, measured peculiar velocity field data, and the covariance matrix ready. Our purpose is to fit the amplitude of matter fluctuation parameter $\sigma_{8}$ and the combined parameter $\Omega^{0.6}_{\rm m}h$ with
the pairwise velocity data. We formulate a log-likelihood function as
\begin{eqnarray}
 -\log \mathcal{L}(\mathbf{\theta}) & = & \sum_{ij} \left(v^{\rm t}_{12}(r_{i}; \mathbf{\theta})- v^{\rm d}_{12}(r_{i}) \right)C^{-1}_{ij} \nonumber \\
& \times & \left( v^{\rm t}_{12}(r_{j}; \mathbf{\theta})- v^{\rm d}_{12}(r_{j}) \right), \label{eq:chi2-final}
\end{eqnarray}
where the indexes ``t'' and ``d'' mean ``theory'' and ``data'' respectively, and $\mathbf{\theta}$ represents the parameters of interest. Maximizing this likelihood function will give us the estimate of the cosmological parameter $\mathbf{\theta}$.

\section{Results}
\label{sec:results}

\begin{figure}
\centerline{\includegraphics[width=3.3in]{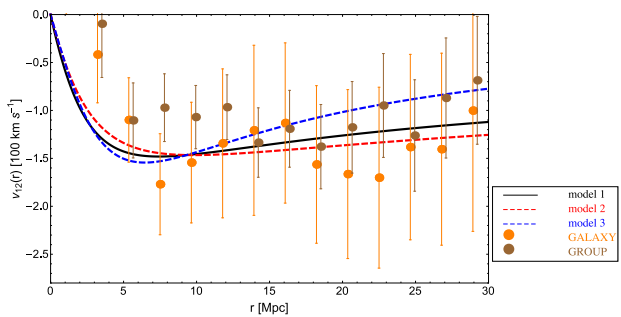}}
\caption{$v_{12}(r)$ for the \galaxy and \group catalogues with the data points calculated from Eq.~(\ref{eq:pairwise-estimate}) and error bars from the square root of Eq.~(\ref{eq:cov-Cij}). The black, red dashed, and blue dashed lines are for Models 1, 2, and 3, respectively, by using {\it Planck} 2015 best-fitting cosmological parameters. The measured $v_{12}(r)$ is separated into 14 bins, and the error bars between each bin are highly correlated.} \label{fig:v12fig}
\end{figure}


\begin{figure*}
\leftline{\includegraphics[width=1.9in]{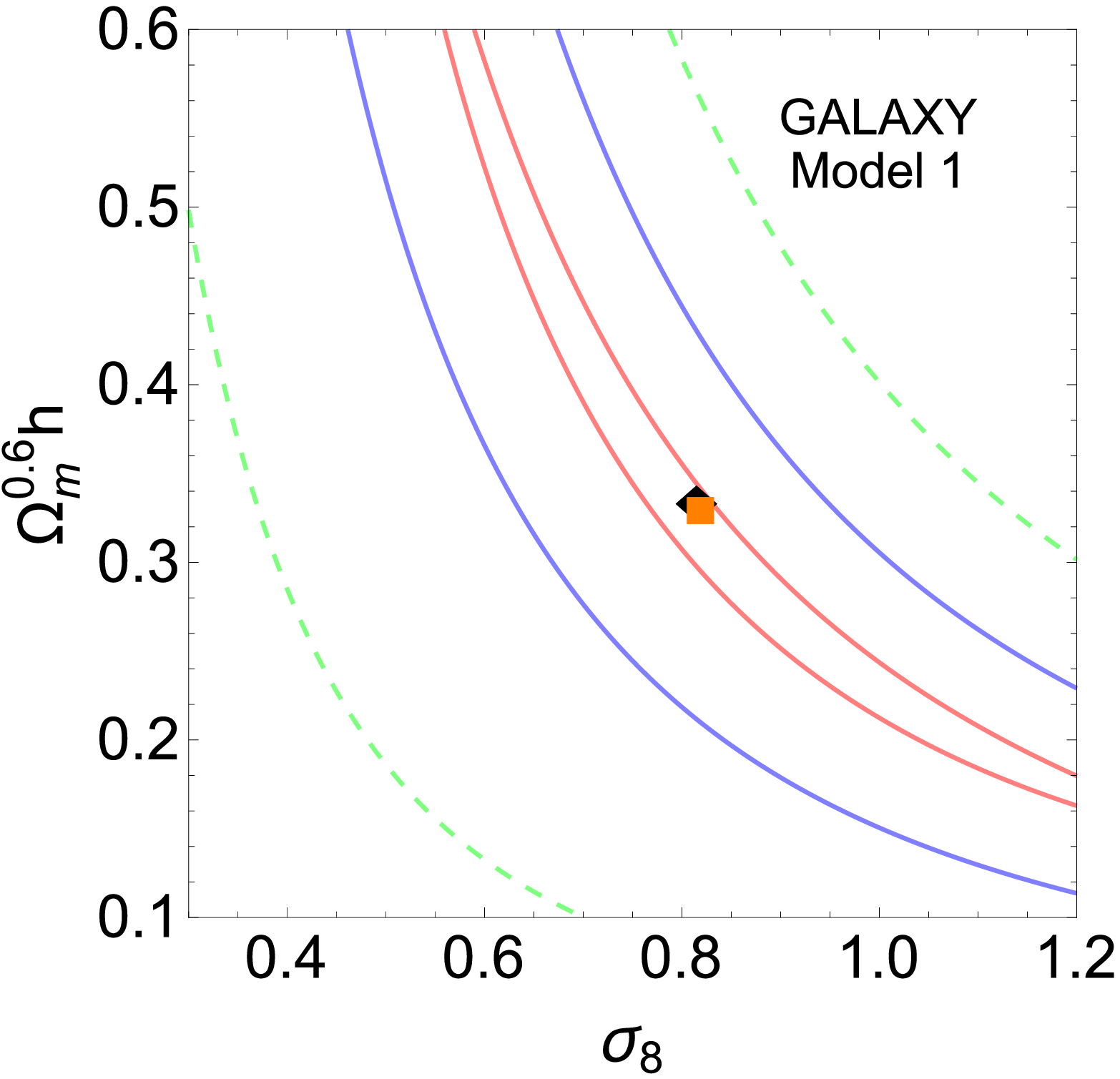}
\includegraphics[width=1.9in]{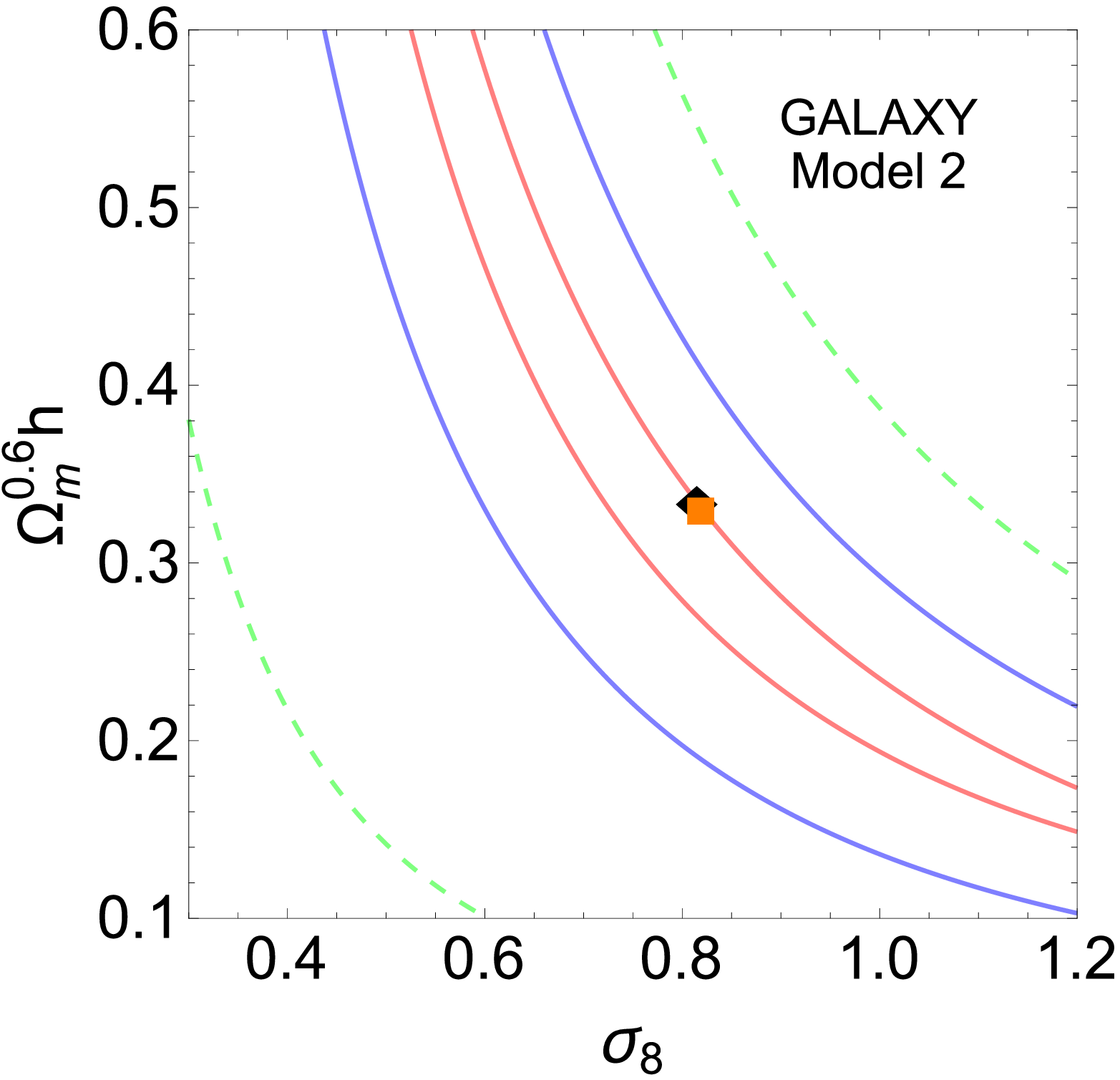}
\includegraphics[width=1.9in]{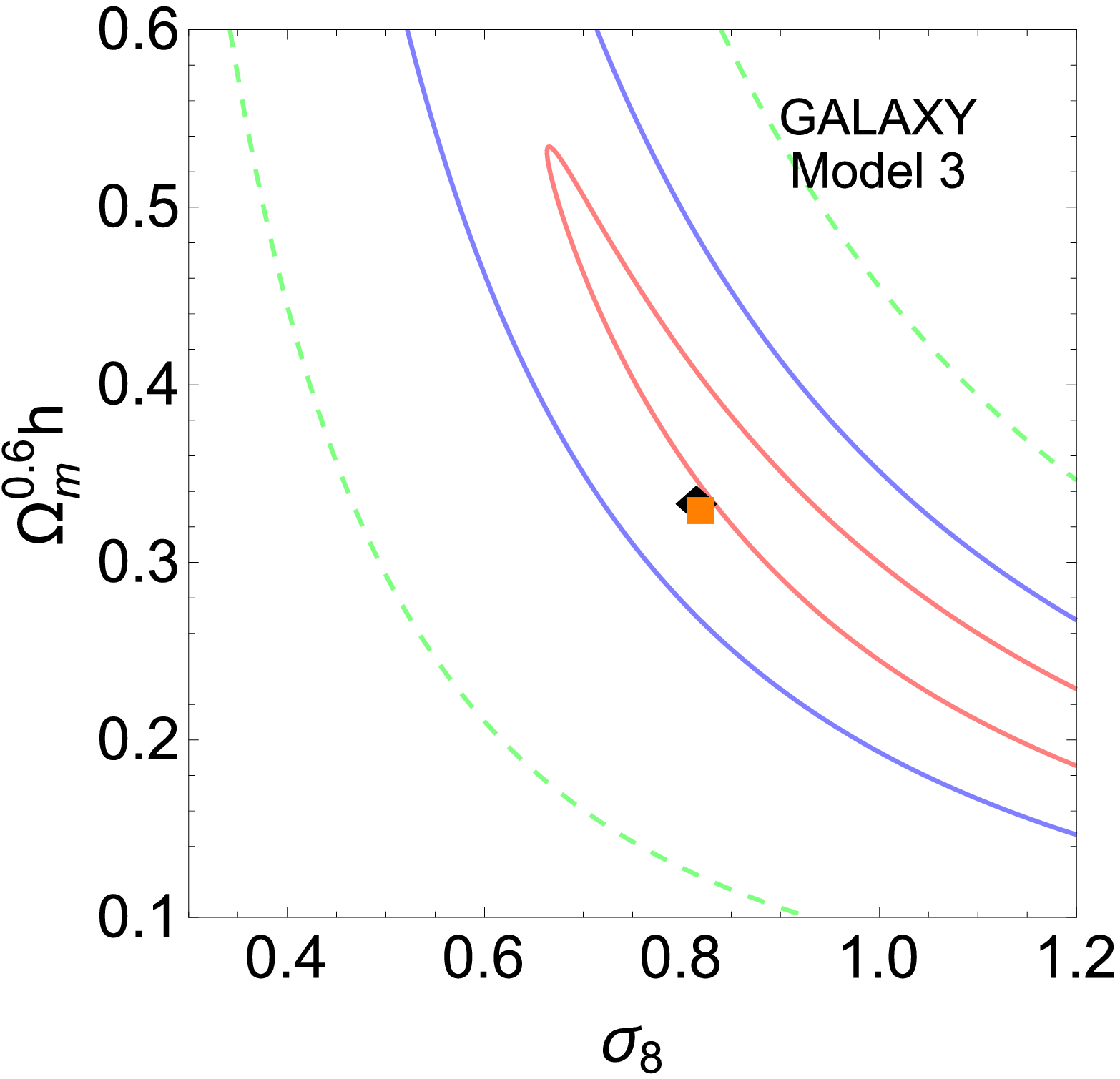}}
\leftline{\includegraphics[width=1.9in]{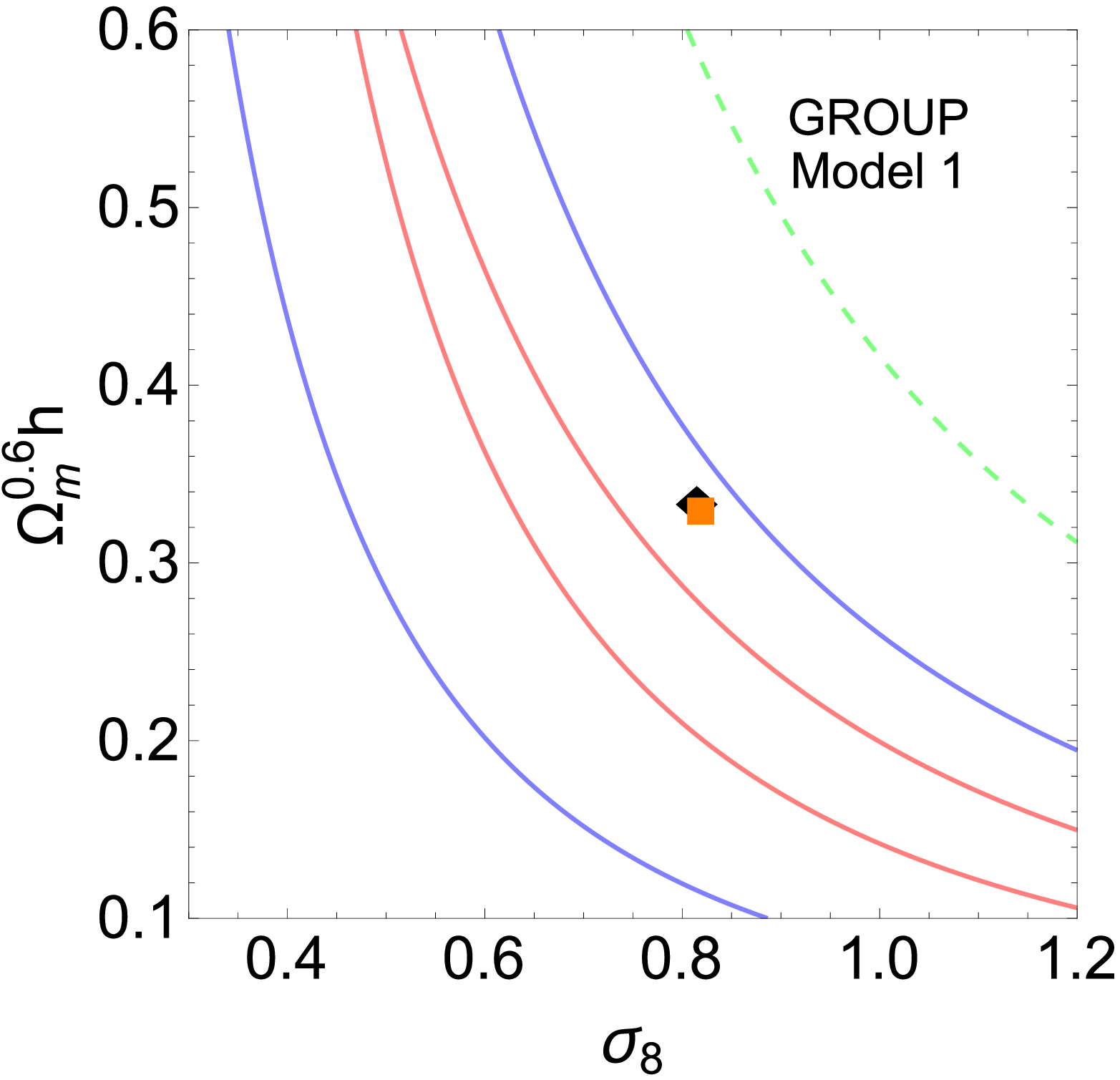}
\includegraphics[width=1.9in]{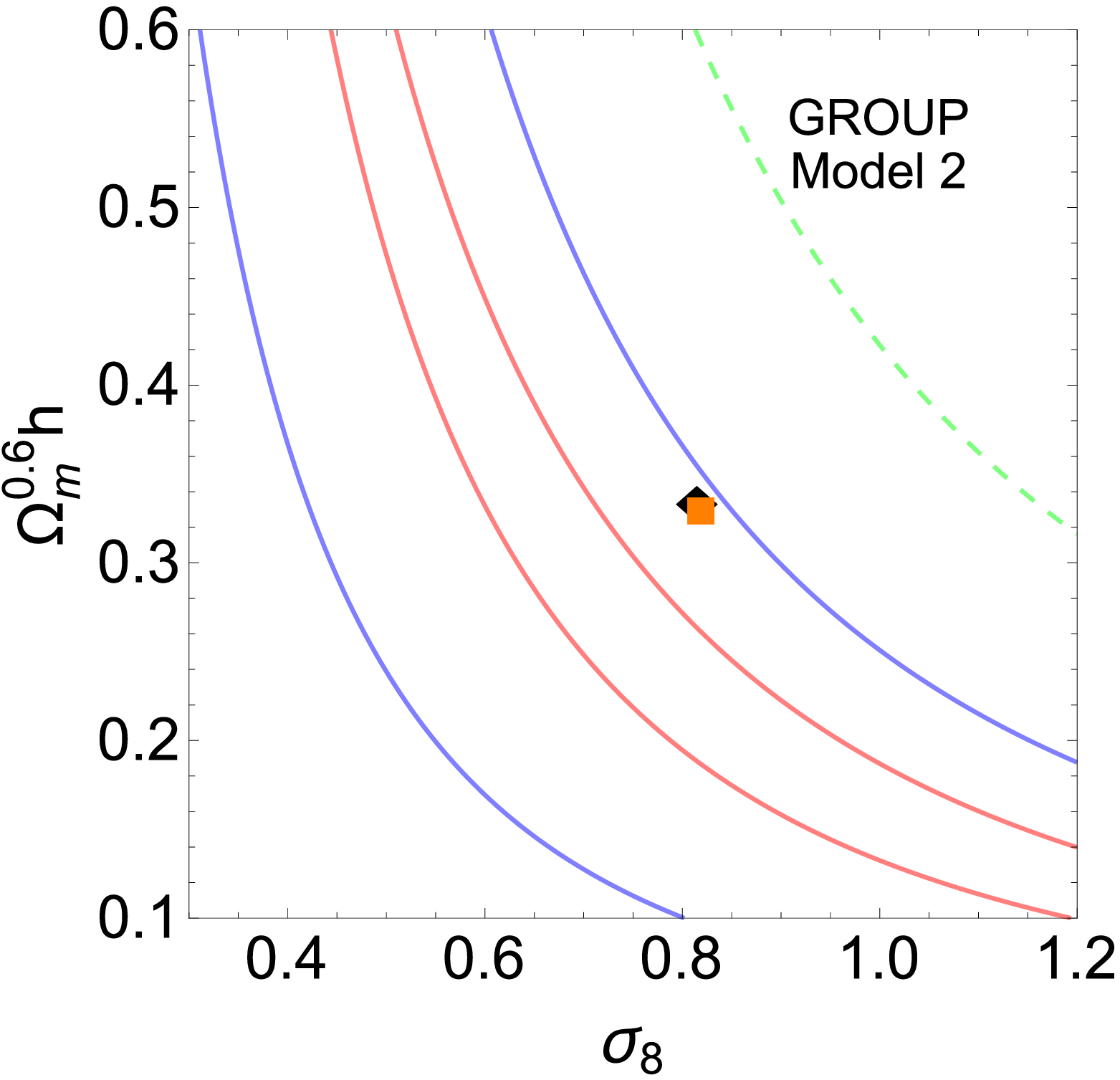}
\includegraphics[width=2.7in]{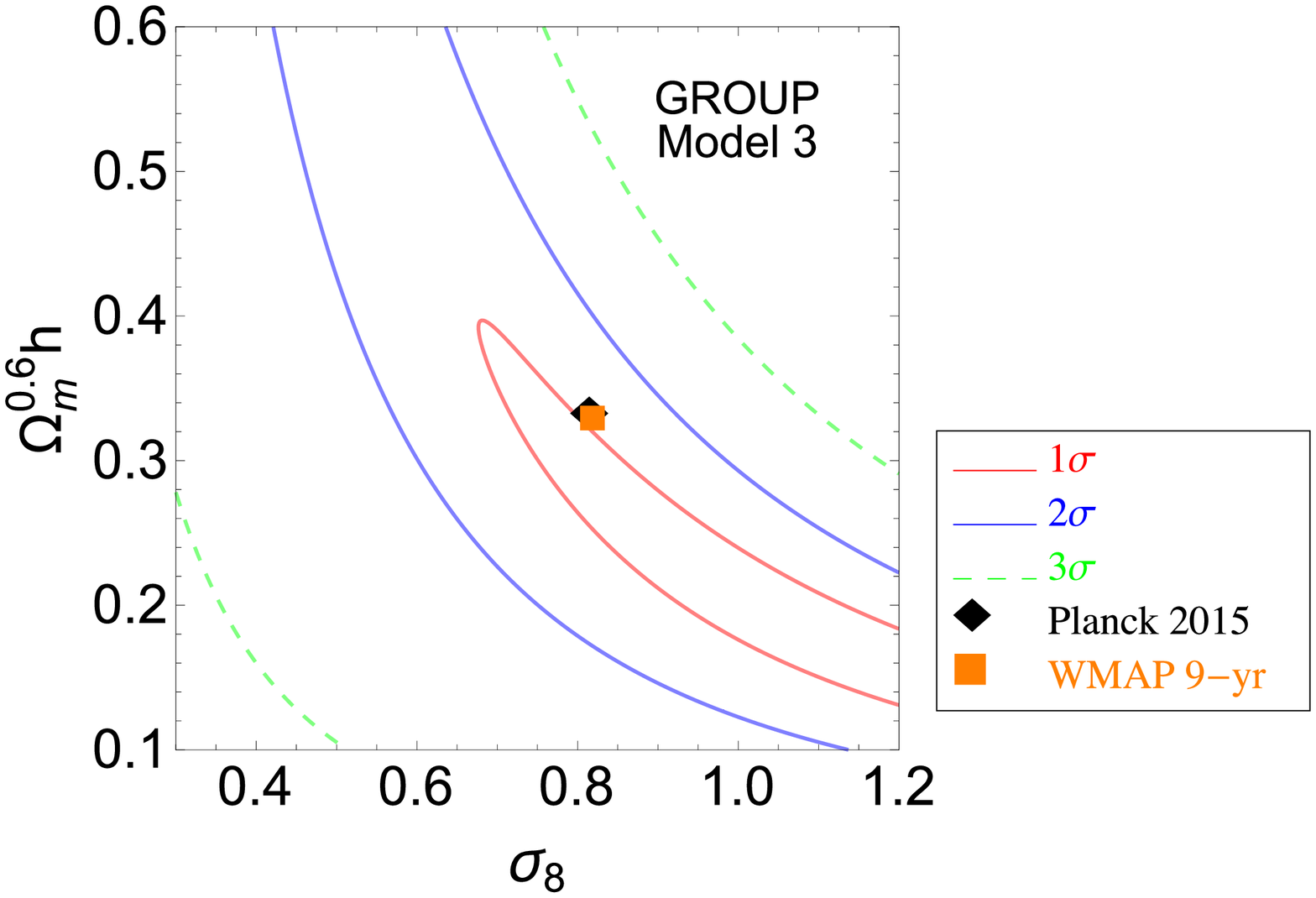}}
\caption{Joint constraints on \omgmh --$\sigma_{8}$ parameters with \galaxy and \group catalogues for three different models. Red, blue, and green dashed lines are the $1$, $2$, $3$ $\sigma$ confidence levels, respectively. The best-fitting cosmological parameters from \wmap 9-year~\citep{Hinshaw13} and \planck 2015 results~\citep{Planck2015-13} are \{$\Omega_{\rm m}$, $\sigma_{8}$, $h$\}=\{0.272, 0.82, 0.704\}, and \{0.309, 0.816, 0.677\}, respectively. The {\it WMAP} 9-year values are taken as the {\it WMAP}9+eCMB+BAO+$H_{0}$, and {\it Planck} 2015 parameter values are taken as {\it Planck} TT+TE+EE+lowP+lensing+ext, where ``lowP'' means low-$\ell$ polarization data and ``ext'' means BAO+JLA+$H_{0}$.}
 \label{fig:likelihood}
\end{figure*}

In Fig.~\ref{fig:v12fig}, the three models are plotted with {\it Planck} 2015 best-fitting cosmological parameters, which seems to be consistent with both \galaxy and \group data sets. However, the error bars in Fig.~\ref{fig:v12fig} are highly correlated (Fig.~\ref{fig:covfig}), so we need to use the full covariance matrix to derive more quantitative results.

We now present our final results for the likelihood analysis. In Fig.~\ref{fig:likelihood}, we plot the joint constraints on \omgmh and $\sigma_{8}$ from \galaxy and \group catalogues, for three theoretical models. The black-diamond and orange-square marks are the best-fitting parameters from the {\it WMAP} nine-year results and {\it Planck} 2015 results. One can see that the joint constraints from the \galaxy catalogue (upper panels) contain the best-fitting point of the cosmic microwave background (CMB) survey at around $1\sigma$ CL, which indicates that they are highly consistent with the $\Lambda$CDM model with {\it WMAP} or {\it Planck} cosmological parameters. For the \group catalogue (lower panels), although the best-fitting parameter values from CMB are not within the $1\sigma$ CL of the joint constraints, they are consistent with  pairwise velocity constraints at $2\sigma$ CL, which is not big enough to claim any discrepancy. In addition, as shown in~\citet{Valentino15}, if one releasing 12 cosmological parameters in {\it Planck} likelihood code, the $\sigma_{8}$--$\Omega_{\rm m}$ tension could be resolved.

Comparing the six panels, we conclude that the current tightest constraint on cosmological parameters is from the \group catalogue with parameter values as
\begin{eqnarray}
\Omega^{0.6}_{\rm m}h = 0.102^{+0.384}_{-0.044},\, \sigma_{8}=0.39^{+0.73}_{-0.1} \,\, ({\rm for\, Model \, 1}). \label{eq:values}
\end{eqnarray}

Looking into the future, the potential for enhancing the constraining power of the pairwise velocity field is embodied in the following aspects.
\begin{enumerate}

\item Better modelling of the pairwise velocity from simulations of the large-scale structure. By developing a numerical simulation, \citet{Slosar06} investigated the distribution of velocities of pairs at a given separation taking both one-halo and two-halo contributions into account. Later on, \citet{Lam11} studied how primordial non-Gaussianity affects the pairwise velocity probability density function by using an analytical model and the N-body simulations. More recently, \citet{Thompson12} have investigated the velocity distribution of dark matter halo pairs in large N-body simulations ($250\,h^{-1}$Mpc--$1\,h^{-1}$Gpc) and examined the pairwise halo velocities with high velocity bullet cluster samples. These efforts at more accurate modelling of pairwise velocity field will continue and lead to the closer description of the observed pairwise velocity data.

\item Better calibration of the distance estimate and the reduction of distance error. The most important aspect of enhancing cosmological constraints on a pairwise velocity field is to improve the distance estimate. Currently, the Tully-Fisher~\citep{Tully77} and FP~\citep{Djorgovski87,Campbell14} relations lead to the distance estimate in the range of $25$--$30$ per cent. For instance, for the FP distance estimator, the total scattering of distance $r$ constitutes intrinsic scatters ($\sim 20$ per cent), FP slope multiplied the observational error ($\sim 18$ per cent) and the photometric error ($\sim 3$ per cent). The bulk part, i.e. intrinsic scatters, is due in part to the effect of stellar population age variations on $M/L$, but the very large uncertainties on individual age estimates mean that it is very difficult to correct the effect~\citep{Campbell14}. Therefore, much effort needs to be devoted to investigating those effects that strongly affect intrinsic scattering.

\item Denser and broader sky coverage. These two factors in total will lead to providing more samples for each individual distance bin so that the {\it rms} noise level will be reduced as the sample size increases. In addition, since the Tully-Fisher survey needs to measure the HI line width, they are limited by the flux of HI measurement~\citep{Masters06,Catinella07}. The Fundamental Plane survey measures the galaxy's surface brightness, so it is limited by the sensitivity of the telescope (telescope size, exposure time, atmospheric interference, etc.). For these reasons, the faint galaxies are hard to measure, and the density in the survey volume is limited. As a rough estimate of the current state-of-the-art 6dF galaxy survey, the sampling density at $z_{1/2}\simeq 0.05$ is $\rho=4 \times 10^{-3}h^{3}$Mpc$^{-3}$~\citep{Campbell14}.

\item Deeper redshift and distance surveys. The survey volume is an important factor in improving the dark energy and modified gravity constraints. This is not only because the deeper survey provides more samples, but also because it provides estimates of the growth of the structure at different redshifts. As a result, for a much deeper survey, one can separate the pairwise velocity samples into different redshift bins and constrain the $f(z)\sigma_{8}$ value in Eq.~(\ref{vel_rnrm1}). The redshift evolution of $f(z)\sigma_{8}$ in the regime of $z=0$--$1$ constitutes a sensitive probe of the modified gravity and dark energy, which has been demonstrated in~\citet{Hudson12} and \citet{Planck2015-13}.

\item Alternative velocity probe of the kinetic Sunyaev-Zeldovich effect~\citep{Sunyaev72,Sunyaev80}. The kinetic Sunyaev-Zeldovich effect (kSZ) is the secondary anisotropy of the CMB produced by the inverse Compton scattering of the moving electrons. The temperature anisotropy of the CMB is therefore proportional to the line-of-sight peculiar velocities of the electrons, i.e. $\Delta T/T=-(\sigma_{\rm T}/c)\int \der l \,n_{\rm e}(r)(\mathbf{v}\cdot\mathbf{\hat{n}})$. Thus, the accurate measurement of the kSZ effect from CMB observations and the precise modelling of intergalactic gas can, in principle, lead to determining the pairwise velocity field without suffering from systematics from optical survey. The pairwise momentum of the temperature field in~\citet{Hand12} gave the first detection of the kSZ effect by using the ACT data. More recently, \citet{Planck2015-kSZ} have re-measured the pairwise momentum field by using {\it Planck} 2015 data and cross-correlated the temperature field with the reconstructed peculiar velocity field. For the first time, they find the $3$--$3.7\sigma$ CL detection of this correlation, which indicates that the gas is extended much beyond the virial radii of dark matter halo, constituting direct evidence of the low-density, diffuse baryons~\citep{Planck2015-kSZ,Carlos15}. The effort of accurately measuring kSZ effect will carry on and will improve the peculiar velocity field measurement from the channel different from optical surveys.

\end{enumerate}

\section{Conclusion}
\label{sec:conclude}

In this work, we have developed a Bayesian statistics method for using a galaxy pairwise velocity field to estimate cosmological parameters. We first reviewed the pairwise velocity estimator developed in \citet{Ferreira99} to calculate the relative motion between galaxy pairs with a separation distance between $0$ and $30\hmpc$, and then we review three theoretical models of the mean pairwise velocity field, especially the one used in \citet{Juszkiewicz99}, which was calibrated against N-body simulation.

We then focused our effort on simulating mock galaxy catalogues and calculated the log-likelihood function of the pairwise velocity field. We first calculated the covariance matrix between any two line-of-sight velocities in a given survey and simulated the $10^{3}$ mock catalogue according to this covariance matrix. Then we calculated the pairwise velocity field for each of these mock catalogues and calculated the variance of the $v_{12}(r)$. In this way, we obtained a covariance matrix of the $v_{12}(r)$ field and used this in the likelihood to estimate cosmological parameters $\Omega^{0.6}_{\rm m}h$ and $\sigma_{8}$. By using the \cf2 catalogue, which consists of over $8\ 000$ galaxies over the full sky out to $25\ 000\kms$, we show that the results are consistent with the {\it WMAP} nine-year data and {\it Planck} 2015 data, and the error bars are currently too big to draw any concrete conclusion on modified gravity models or dynamical dark energy\footnote{In this sense, we think what is claimed in~\citet{Hellwing14} is too optimistic.}. We hope therefore that future data with more samples on the local Universe may help to pin down the sample variance and improve pairwise velocity field statistics.

\vskip 0.1 truein

\noindent \textit{Acknowledgments:} We would like to thank Chris Blake, Helene Courtois, Elisabete da Cunha, Tamara Davis, Hume Feldman, Pedro Ferreira, Andrew Johnson, and Jeremy Mould for helpful discussions. We also acknowledge the use of the ``VizieR'' astronomy online database and {\sc camb} package. Y.Z.M. acknowledges support from an ERC Starting Grant (No. 307209), and P.H. acknowledges support by the National Science Foundation of China (No. 11273013) and by the Open Project Program of State Key Laboratory of Theoretical Physics, Institute of Theoretical Physics, Chinese Academy of Sciences, China (No. Y4KF121CJ1).

\appendix
\section{Simulating a catalogue given its covariance matrix}
\label{sec:simulate}
The $G_{ij}$ is a symmetric, positive-definite matrix, where we can use the singular value decomposition (SVD) method to simulate the velocity vector $v_{i}$ ($i=1,...,N_{\rm data}$). The $G$ matrix can be decomposed into
\begin{eqnarray}
G=U\cdot S \cdot U^{\rm T}, \label{eq:svd1}
\end{eqnarray}
where $U$ is the left matrix, and its transpose is the right matrix, and the $S$ matrix is the diagonal matrix with $N_{\rm data}$ element that contains the eigenvalues of matrix $G$. Since $G$ is a positive-definite matrix, the elements in $S$ should be positive. Then we can simulate an array of Gaussian random variables $\tilde{v}$ with zero mean and variance $S_{i}$ ($i=1$, $2$, ..., $N_{\rm data}$) to satisfy
\begin{eqnarray}
\langle \tilde{v}\tilde{v}^{\rm T} \rangle = S.
\end{eqnarray}
Finally, the real velocity vector we want to obtain is $v=U\tilde{v}$, because
\begin{eqnarray}
\langle v v^{\rm T} \rangle &=& \langle U\cdot\tilde{v} \tilde{v}^{\rm T} \cdot U^{\rm T} \rangle \nonumber \\
&=&  U\cdot \langle \tilde{v} \tilde{v}^{\rm T} \rangle \cdot U^{\rm T}  \nonumber \\
&=& U\cdot S \cdot U^{\rm T} \nonumber \\
&=& G;
\end{eqnarray}
i.e. the covariance matrix of $v$ is $G$. In this way, we simulate a line-of-sight velocity vector with dimension $N_{\rm data}$, which satisfies the desired covariance matrix.

%

\end{document}